\documentclass[12pt]{iopart}
\usepackage{graphicx}
\usepackage{tabularx}
\usepackage{multirow}
\usepackage{tablefootnote}
\usepackage{hyperref}
\usepackage{xcolor}
\usepackage{footmisc}

\newcommand{\erf}{\textrm{erf}}

\begin{document}

\title[New Approach for Fuzzy-Classification of Gravitational-Wave Signals]{A Novel Multi-Layer Modular Approach for Real-Time Fuzzy-Identification of Gravitational-Wave Signals}

\author{Francesco Pio Barone\footnote{The author F.P. Barone was previously with University of Catania and is now associated with University of Padua, Padua, Italy.}}
\address{Dipartimento di Fisica e Astronomia, University of Catania}
\author{Daniele Dell'Aquila\footnote{Previously at Dipartimento di Scienze Chimiche, Fisiche, Matematiche e Naturali, University of Sassari, Sassari, Italy.}}
\address{Dipartimento di Fisica ``Ettore Pancini", University of Naples ``Federico II", Napoli, Italy}
\address{INFN-Laboratori Nazionali del Sud, Catania, Italy}
\author{Marco Russo}
\address{Dipartimento di Fisica e Astronomia, University of Catania}
\address{INFN-Sezione di Catania, Catania, Italy}
\ead{daniele.dellaquila@unina.it}
\vspace{10pt}
\begin{indented}
\item[]June 2023
\end{indented}

\begin{abstract}
Advanced LIGO and Advanced Virgo ground-based interferometers are instruments capable to detect gravitational wave signals exploiting advanced laser interferometry techniques. The underlying data analysis task consists in identifying specific patterns in noisy timeseries, but it is made extremely complex by the incredibly small amplitude of the target signals. In this scenario, the development of effective gravitational wave detection algorithms is crucial. We propose a novel layered framework for real-time detection of gravitational waves inspired by speech processing techniques and, in the present implementation, based on a state-of-the-art machine learning approach involving a hybridization of genetic programming and neural networks. The key aspects of the newly proposed framework are: the well structured, layered approach, and the low computational complexity. The paper describes the basic concepts of the framework and the derivation of the first three layers. Even if, in the present implementation, the layers are based on models derived using a machine learning approach, the proposed layered structure has a universal nature. Compared to more complex approaches, such as convolutional neural networks, which comprise a parameter set of several tens of MB and were tested exclusively for fixed length data samples, our framework has lower accuracy (e.g., it identifies $45\%$ of low signal-to-noise-ration gravitational wave signals, against $65\%$ of the state-of-the-art, at a false alarm probability of $10^{-2}$), but has a much lower computational complexity (it exploits only $4$ numerical features in the present implementation) and a higher degree of modularity. Furthermore, the exploitation of short-term features makes the results of the new framework virtually independent against time-position of gravitational wave signals, simplifying its future exploitation in real-time multi-layer pipelines for gravitational-wave detection with new generation interferometers.
\end{abstract}

%
\vspace{2pc}

\section{\label{sec:intro}Introduction}
During the last decade, the Advanced Laser Interferometer Gravitational-Wave Observatory (aLIGO) \cite{Aasi15} has set in motion the era of gravitational wave astrophysics through the first direct detection of a gravitational wave \cite{Abbott16d}\footnote{The first direct observation of a gravitational wave was announced on February 11th, 2016, jointly by the LIGO Scientific Collaboration and Virgo Collaboration.}. Gravitational waves (GWs) have now become a new, key, means to understand our Universe, leading to a series of fundamental discoveries. For example, the detection of GW150914 made it possible, for the first time, to observe a binary black hole (BBH) merger \cite{Abbott16, Abbott16b, Abbott16e}, i.e. the coalescence of two massive black holes, an extremely catastrophic astrophysical event. This gave, for the first time, the unique opportunity to quantitatively test the predictions of Albert Einstein's general theory of relativity for the motion of a compact-object binary in the large-velocity, highly-nonlinear regime \cite{Abbott16b,Abbott16c,Li12,Mishra10}.

Alongside the two aLIGO interferometers, by the end of the second aLIGO discovery campaign (O2), in 2017, the European Advanced Virgo (aVirgo) detector \cite{Acernese15} took part in the GW detection network. The availability of an international, second generation, three-detector network had a crucial role in the detection, with improved sky location, of numerous GW events which were found to be consistent with the coalescence of BBH mergers (see e.g. GW170814, GW170823, GW171014, GW190521 \cite{Abbott17c,Abbott19,Abbott20}).

Besides the identification of GW events confidentially interpreted as due to BBH mergers, GW170817 is the first GW detection linked to the coalescence of two neutron stars (BNS) \cite{Abbott17}. This gravitational wave event was detected alongside with its electromagnetic counterpart, GRB170817A \cite{Abbott17d,Abbott17e}, i.e. an electromagnetic emission that accompanies the BNS coalescence, leading to the first direct confirmation that neutron star mergers are the progenitors of gamma-ray bursts. This unique joint detection has inaugurated the era of multi-messenger astronomy, a strongly multi-disciplinary research field at the interface of experimental and theoretical physics.

The second generation of GW detectors is now poised to significantly enhance the sensitivity of the observations to a much larger search volume \cite{Abbott16f} and to new GW sources for which the expected signal cannot be modeled with our present theoretical understanding. In addition, one of the major goals of the community is that of enabling future multi-messenger observations with the joint effort of other astronomical facilities \cite{Abbott16g,Abdo13,Aartsen20,Abe18,AdrianMartinez07}. Detecting a multi-messenger event requires the combined effort of GW interferometers and other kind of astronomical detectors, which, ideally, should be oriented towards the sky location of the GW signal in a relatively short time after the GW event is detected. In this framework, the development of new, fast, methods to detect GW signals is crucial, not only to enable the observation of multi-messenger events, but also to help deepen the parameter space of astrophysical searches \cite{Indik18}.

There are two main classes of algorithms conventionally used to search for GW transients in ground-based interferometers: matched-filtering algorithms \cite{Owen99,Usman16,Adams16} and coincident excess-power searches \cite{Klimenko08,Klimenko16,Sylvestre02,Chatterji04}. The first class of algorithms usually relies on a number of previously modeled waveforms, representing the expected GW signals for various astrophysical sources, which constitute a so called \emph{template bank}. They are particularly effective to detect well-modeled GW signals, but have poor interpolation capabilities between templates and cannot efficiently extrapolate outside of the parameter space covered by the \emph{template bank}. Furthermore, they are typically computationally costly, and their computational complexity increases with the number of templates to test. For this reasons, there are ongoing efforts to try to reduce the size of template banks, in order to accelerate the offline parameter estimation via matched-filtering \cite{Indik18}. The second class of methods allows instead to identify even unmodeled or weakly modeled waveforms, as they are often treated as free parameters of a likelihood formalism \cite{Klimenko16}.

It has been recently proven that state-of-the-art deep learning artificial intelligence techniques can help to facilitate the search for GWs with second-generation interferometers \cite{George18b,Gabbard18,Razzano18}. In particular, novel convolutional neural networks (CNNs) turned out to be ideal to model matched-filtering algorithms in the search of inspiral-merger-ringdown BBH waveforms \cite{George18b,George18,Gabbard18}. This class of algorithms is promising as it allows to significantly reduce the size of template banks while emulating the performance of matched-filtering searches \cite{Gabbard18}. Compared to matched-filtering, CNNs can interpolate between the templates used to derive the model and have more robust generalization capabilities \cite{George18}. Unfortunately, their results strongly depend on the time position of the waveform within the analyzed time window, making challenging their use in real applications characterized by a continuous data streams. Given that machine learning (ML) techniques are extremely promising to help the detection of GWs in state-of-the-art experiments, it is interesting to try to derive new algorithms, alternative (and complementary) to those based on CNNs, more suitable to be used in continuous data stream applications and more robust against the completeness of the template banks. In addition, in our opinion, the modularity of the approach is also an element to be seriously taken into account. As well known in other fields, such as computer internetworking and operative systems, modularity simplifies research and development and makes it possible to more easily improve the performance of the system with even continuous advancements.

In this paper, we introduce a novel ML approach for real-time fuzzy-identification of GW signals in second generation ground-based interferometers that exploits, for the first time in this research field, a short-term, layered, approach inspired by speech processing techniques. The paper describes the basic concepts of the framework and proposes a possible low-complexity \emph{ground model}, which is ideal to be used in the first layer of our approach. We also describe the detailed derivation of two possible models for the interpretation of the results of the first layer, which serve to build two possible versions of the second layer, and discuss the possible implementation of a third (\emph{final}) layer, which would issue the interferometer alerts.

In the newly proposed approach, data segments from GW detectors are suitably subdivided into smaller, overlapped, frames, where numerical features derived from speech processing are calculated. ML techniques can be then used to derive a model, exploited by the first layer of the approach, capable to link numerical features to an output that represents the fuzzy-degree of belonging of the frame to the GW class. Each successive layer exploits the output of the previous layer and a dedicated model for its analysis.

To derive the ground model presented here, we adopted state-of-the-art symbolic regression techniques \cite{Russo16,Russo20}. In this way, the derived model has an analytical formulation and is particularly suitable for highly-optimized implementations. The underlying methods rely on a hybridization of evolutionary computing and Neural Networks (NNs). Such techniques are well-developed in other research fields (see e.g. Refs.~\cite{Campobello20,Russo14,Buccheri21,Buccheri22,DellAquila21}) but were not previously proposed for GW searches. The  model described in this paper, and used in the first layer, is trained and tested exploiting a bank of simulated BBH waveforms \cite{Husa16,Khan16} in synthetic Gaussian noise, produced according to the aLIGO design sensitivity curve\footnote{We adopt the LIGO-P1200087 design sensitivity curve.} previously used in \cite{Gabbard18}. The choice of assuming stationary the instrumental noise is a simplified hypothesis to facilitate the comparison with previous ML approaches published in the literature \cite{Gabbard18,George18,George18b}. We demonstrate that our framework, even using the simple ground model described in the paper, can achieve a similar performance, in identifying GW signals, than other tools based exclusively on CNNs, but with some key advantages: (1) no hypothesis is made on the time position of the waveform in the stream, facilitating the future implementation of the framework in continuous searches; (2) with the proposed first and second layers, the framework relies only on an extremely reduced number of numerical features, making it suitable for highly-optimized implementations in real-time applications; (3) the approach is layered, allowing a larger degree of versatility and making the framework ideal to be integrated even in existing pipelines for online analysis.

The paper is organized as follows, Section~\ref{sec:framework} contains a general description of the proposed framework and symbolism; Section~\ref{sec:ground_model} describes the derivation of the ground model (used in the first level of the framework), including assumptions, dataset, numerical features and the artificial intelligence approach used to derive the model, in addition, we also provide its analytical formulation; in Section~\ref{sec:results}, we describe the second level of the framework, which is to analyze the results of the ground model, and compare the accuracy of our approach against that obtained by state-of-the-art models. Finally, Section~\ref{sec:conclusions} is to summarize our findings and discuss their impact on the design of future pipelines for real-time analysis of GWs, including a possible implementation of a third level.


\section{\label{sec:framework}The framework: general overview, symbols, approach}
Gravitational waves are space-time ripples that can be detected monitoring modulations in the length of the arms of an interferometer with state-of-the-art laser interferometry techniques. Interferometer data are timeseries usually characterized by a loud and persistent background noise, which makes particularly challenging the identification of candidate GW signals, extremely weak signals compared to the background. The ultimate goal of the newly proposed framework is to develop a method capable to perform a real-time analysis of a continuous stream of interferometer data in a versatile and computationally convenient way, for gravitational wave applications.

Let us indicate with \emph{local analysis} a GW search restricted to a certain frame, of fixed length, of the data stream. This could be done through a suitable model, trained, for example with a ML approach, to determine whether or not a GW signal is present within the given frame, i.e. to determine to what fuzzy degree a given frame represents a GW signal. An extension from a local to a \emph{continuous analysis}, which is more realistic for a real interferometer, usually involves exploiting the local results on several moving windows \cite{George18}. Local classification is therefore a good starting point for the successive continuous implementation using moving windows. Previous applications with ML-based techniques \cite{Gabbard18,George18,George18b} have mainly focused to \emph{locally} distinguish noise-only segments from data segments containing a gravitational wave signal, assuming that only stationary noise is present. The derived models had outstanding classification capabilities in local analysis cases, given that the position of the maximum of the BBH waveform amplitude is well-placed in a given fractional range of the window interval, usually with a width of the order of $20\%$ of its entire length. For this reason, it is difficult to predict the outcome of these models in a real use case, where one has a continuous timeseries analyzed in moving windows, in which the peak amplitude is often outside of such fractional range.

\begin{figure}
	\centering
	\includegraphics[width=0.60\textwidth]{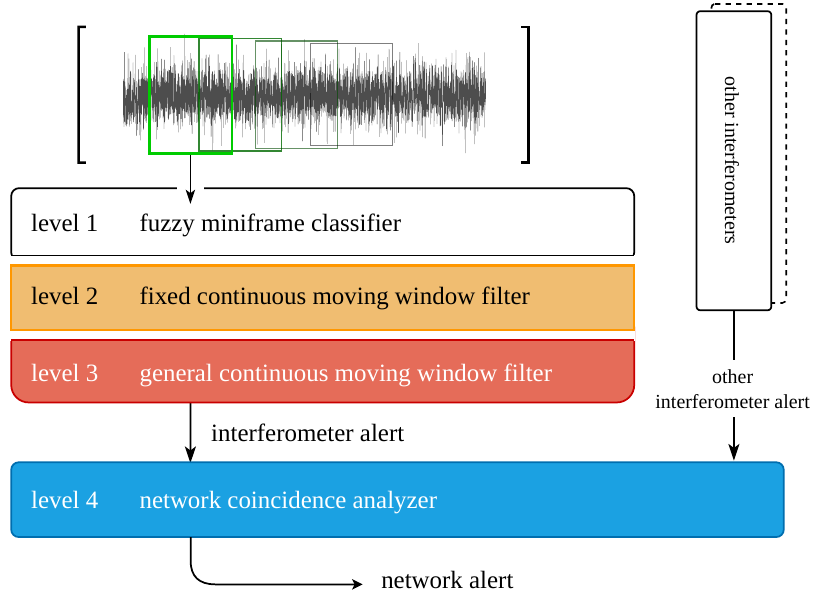}
	\caption{\label{pic:framework_layout} The basic structure of the multi-layer framework proposed in this paper. Levels 1-3 perform an analysis of the timeseries starting from short-term features inspired by speech processing. Level 3 issues an interferometer alert. The latter can be fed into a hypothetical level 4, which would identify coincident alerts in an interferometer network, issuing a global \emph{network alert}.}
\end{figure}
The concept proposed in this paper is a multi-layer framework, where the first layer relies on short-term features, extracted from the interferometer timeseries, and each successive layer refines the information of the previous layer. Compared to previously published CNNs models, the use of short-term information, other than complex vectors of raw data, allows to avoid assumptions on the peak amplitude position even for local analysis cases.

The layered structure of the proposed framework is inspired by a paradigm typically used in the engineering of communication and/or operative systems. Layering, for example, provides a distinct advantage in an operating system. The inner structure of each layer can be defined separately, given that the i/o is fixed, and layers can interact with each other as required. This is crucial to more easily create, maintain and update the system. Performing a change in one layer, for example an improvement of its performance, does not affect the lower lying layers, and, sometimes, the successive layers or both. In a similar framework, each layer \emph{virtualizes} the underlying level and gives more refined information to the successive level.

Our framework is composed by several distinct layers, as schematically described in Figure~\ref{pic:framework_layout}. In our solution, if a given layer is changed, only the successive (upper) levels may be affected. Level 1 performs the extraction of short term features and utilizes a previously trained model to predict the content of each short data frame in which the timeseries is analyzed. For each set of short-term features, related to a given data frame, an output value is thus produced and passed to the second layer. Level 2 analyzes a fixed number of consecutive output values from level 1 through a suitable interpretation model. This layer, in turn, performs the local classification of data segments, similarly to what CNN models do starting directly from the raw timeseries in the entire moving window. The output of level 2 is a vector of moving window predictions, which is finally analyzed by level 3 to issue the \emph{inteferometer alert} if a gravitational wave is detected\footnote{When a gravitational wave is detected, an alert is usually issued by the collaboration to inform the scientific community. This helps the search for multi-messenger events.}. This does not account for coincident information from other interferometers. The error (\emph{false alarm}) rate could be drastically reduced by a hypothetical level 4, which would collect level 3 output for an interferometer network and search for coincident alerts, thus issuing a definitive \emph{network alert}.

In this paper, we mainly focus on the detailed implementation of the first two layers of the framework, which deal with the local analysis of the timeseries. The resulting classification capabilities, which are only valid for local segments, can be easily compared with CNNs models. We also suggest a possible implementation for the third level, but, for the sake of clarity, we postpone its full description to a, future, dedicated paper. To easily compare the capabilities of our framework with previous CNN approaches, we restrict our analysis to BBH signals and we assume that the interferometer noise is stationary. This is a simplification previously introduced in a number of studies (see e.g. \cite{Gabbard18,George18,George18b} and references therein).

\subsection{Level 1 design}
The first level exploits an approach derived from speech processing techniques \cite{Beritelli97,Beritelli98,Beritelli99}. The very first step is considering a continuous signal in smaller, overlapped, short data frames of predefined length (\emph{miniframes}, hereafter). After preliminary tests, we settled on a optimal miniframe size of $\omega_{f} = 100$ ms, with an overlap of $\sigma_{f}=50$ ms. Each miniframe is directly processed by the first level through a specific model, which we call \emph{ground model}. The ground model is a miniframe \emph{fuzzy} classifier, which is trained to distinguish between noise-only miniframes (null hypothesis, \emph{negative} output of the classifier) and miniframes containing a BBH signal (alternative hypothesis, \emph{positive} output of the classifier), exploiting numerical features calculated from the data of a given miniframe. The ground model is trained to return an output closer to $1$ if a GW is present inside the miniframe, and closer to $0$ if that is a noise-only pattern. A more detailed discussion of the ground model, including all details of its derivation, is provided in Section~\ref{sec:ground_model}.

\subsection{Upper levels design}
The second level of the framework does not have direct access to the original data segment coming from the continuous stream. Instead, the signal is basically translated into a sequence of numerical values provided by the first layer. The stream of output values passed by the first level to the second level requires a dedicated interpretation. In this paper, we propose two possible level 2 layers (modules, in the modular philosophy), based on two possible interpretation models of level 1 output. The separation of the ground model derivation from its interpretation is therefore a \emph{multi-modeling} procedure (see Sect.~\ref{subsec:level2})). A strong edge of such an approach is the absence of an a priory hypothesis about the GW transient position within the data segment.

Level 2 analyzes the sequence returned by the first layer in fixed chunks of $n_f$ frames (local analysis). Its purpose is the search for a GW transient specific pattern in the space of values of the first layer. We call this step \emph{moving window analysis}, because, at each iteration, the $n_f$ miniframes selection moves forward by a defined number of miniframes, usually $\sigma_{mw}=1$. This layer is crucial to the classification because it looks at the signal features in a wider time span, combining information on adjacent frames. The output of level 2 is a stream of fuzzy-like or strong integer predictions: $1$ stands for a GW transient, $0$ for noise. Ultimately, level 3 issues an interferometer GW alert by analyzing, in a similar manner than that used by level 2, the output of the second layer. A possible implementation of level 3 could, for example, identify clusters of ones after thresholding the fuzzy-like values of level 2. It is reasonable that clusters of ones will represent a GW in the input signal, as well clusters of zeros will mark a noise-only sequence in the original timeseries.

The framework described above is an early implementation, but promising in the continuous analysis of data, as it will be shown in Section~\ref{sec:results}. Once trained, each framework level performs low complexity operations, thanks to the simplicity of the underlying models. The ground model used in the proposed implementation leads to a level-2 accuracy slightly lower than existing state-of-the-art CNN approach, such as \cite{George18,Gabbard18}, but it is outstandingly computationally simple. This makes the entire framework particularly suitable to be implemented in real-time, low-latency, searches even on a single machine, thus avoiding parallel and distributed calculations and related complications.

\section{\label{sec:ground_model}Ground model derivation: the underlying level 1 model}

The ground model is the core of the entire framework, as it is a fuzzy classifier of short GW interferometer signals. In our framework, the ground model performs the \emph{ground} predictions on short data frames. The ground model output is a continuous variable that varies in the range $[0,1]$ (see Section \ref{ssec:analytical_model}). As mentioned before, the latter can be interpreted as the \emph{fuzzy degree of belonging}, of the individual miniframe, to the GW signal class, i.e. it is related to how likely the short data frame contains a GW signal. Because the typical features of a GW signal are contained in a succession of miniframes, other than in a single short data segment, in order to produce a trigger, it is required to interpret the output of the ground model over a number of successive miniframes. This is even a more critical task because of the large instrumental noise level. For these reasons, successive layers are required. 

The ground model is initially derived by using data samples of $1$ second length, either noise-only or noise-prevailing GW inspiral samples are used, in a similar way as done, for example, in Refs.~\cite{George18,Gabbard18}. Therefore, each data sample is subdivided in a given number of overlapped miniframes of fixed time length. For each miniframe, a vector of features is extracted and processed by a ML model, which is the ground model of the approach. After interpretation of model output, the final outcome is a prediction over the entire $1$ second segment: as stated before, we associate $1$ to GW positive samples, $0$ to noise-only samples.

The key for a good ground model lies in its low computational cost. The ground model proposed in this paper meets by design this request.

The following subsections describe the derivation of the ground model and its interpretation in a \emph{local} analysis case (level 2). Subsection \ref{ssec:data_segments} describes on the data samples used to train and test the model. Subsection~\ref{sub:bp} presents some details of the tool used to derive the models, the Brain Project. Subsection \ref{ssec:data_features} introduces the numerical features derived from audio analysis literature and used to derive the present version of the ground model. Finally, subsection \ref{ssec:analytical_model} discusses the analytical formulation of the derived model.

\subsection{\label{ssec:data_segments}Detector and data segments}

\begin{figure}
	\centering
	\includegraphics[width=0.5\textwidth]{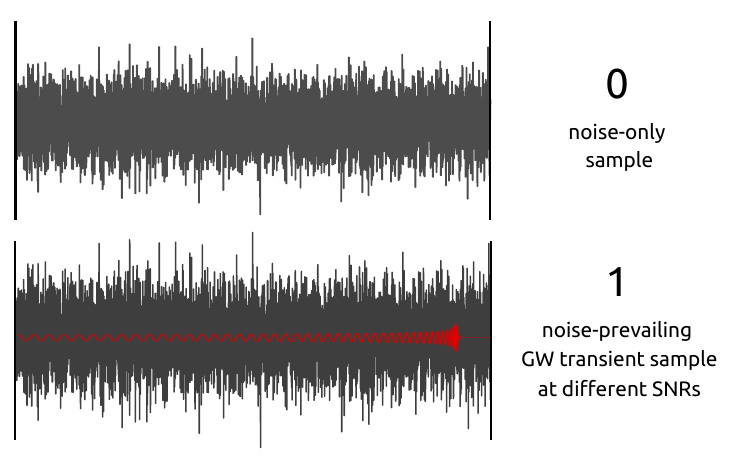}
	\caption{\label{pic:bank} Typical, noise only and signal + noise, whitened data segments used to train the ground model. A noise-only sample is interferometer-like stationary noise generated according to the aLIGO design sensitivity curve. A GW sample is a noisy interferometer signal with a GW transient whose amplitude is scaled to have a target signal-to-noise ratio. The latter requires the generation of a random BBH waveform; in this figure, the red curve (not in scale) marks the BBH waveform as it would appear without the background noise. Each prototype is $1$ second long and sampled at 8192Hz.}
\end{figure}

As previously mentioned, the database used to train and test the model contains simulated interferometer timeseries, either noise-prevailing GW signals or noise-only data samples. The duration of the simulated segments was fixed to $1$ s, as previously done, for example, in Refs.~\cite{Gabbard18,George18,George18b}. This is sufficiently large to contain a typical BBH GW signal (often called \emph{strain}), for which the relevant part that enters the interferometer frequency band lasts usually a few tens of second; for their short duration, BBH signals are usually called \emph{transients}. Figure~\ref{pic:bank} shows a schematic example of time-series used to derive the model. The two key steps required to create GW samples include the generation of a template BBH waveform and its sum to an interferometer noise to match with a target signal-to-noise ratio (SNR). In this way, two types of data segments are produced: \emph{noise only} and \emph{signal + noise} samples.

BBH \emph{template} waveforms, i.e. BBH GW signals according to existing theoretical models, are generated through the LALsuite library \cite{lalsuite}, a package for gravitational and relativistic astrophysics, which is used to perform a numerical simulation of BBH coalescence events (see Appendix~\ref{app:gw}).

Because the simulated source is located at a fixed distance, for a given choice of its astrophysical parameters and a given assumption on the detector noise, signal + noise segments would have exclusively a certain SNR\footnote{For a given set of astrophysically relevant parameters, the amplitude of the GW signal decreases with the distance from the detector at which the source is located. In addition, the source parameters also affect the amplitude and characteristics of the signal.}. However, to derive and test a model capable to detect GWs signals, it is crucial to probe its capabilities at different SNRs. For example, testing the accuracy of a model with different SNRs enables to probe the generalization capabilities of the framework and allows to evaluate the rise in performance with more favorable SNRs. To add detector noise to the time series, data segments containing noise only are generated according to a typical aLIGO design sensitivity curve. Noise was chosen to be exclusively Gaussian. Stationary noise is a simplified hypothesis to facilitate the comparison with previous approaches already published. A number of generated noise segments were instead used to populate the dataset of noise only samples. Finally, before summing the generated GW signals to the stationary noise, the waveform amplitude was suitably scaled to match with the target optimal SNR. The latter is defined according to the following equation \cite{Gabbard18}:
\begin{equation}
\label{eq:rho_opt}
\rho^2_{opt} = \langle u,u \rangle
\end{equation}
where $u$ is the generated template and the inner product $\langle a,b \rangle $ is the noise-weighted cross-correlation defined in Ref.~\cite{Owen99}:
\[\langle a,b\rangle=4 Re\left[\int_{f_{min}}^\infty \frac{\tilde{a}^\star(f)\tilde{b}(f)}{S_n(f)} \right]\]
being $\tilde{a}(f)$ ($\tilde{b}(f)$) the frequency-domain representation of the signal $a$ ($b$) and $S_n$ the single-sided detector noise PSD; $f_{min}$ is the frequency of the gravitational-wave template at the beginning of the time-series.

Before splitting the generated segments into the datasets used to derive the models, the time series are suitably whitened in order to flatten the frequency distribution dominated by the detector noise. This is a common practice in the analysis of GW data. The stationary noise and the smoothness of the simulated aLIGO noise spectrum allowed to use a short, $0.5$ s, padding. To this end, the segments had an additional $0.5$ s of data before and after each generated $1$ s time series. Signal content, after whitening, was truncated using a Tukey window ($\alpha=1/8$) to its central $1$ s of data. In addition, similarly to previous CNNs studies and to allow a meaningful comparison of the local results, the peak amplitude of each waveform within the final time series, was randomly placed within the range [$0.75$ s, $0.95$ s].

For each model derived in this paper (namely the ground model, which serves the proposed level 1, and two interpretation models used in two possible versions of level 2), we build a training, a cross-validation and a testing dataset. The cross-validation dataset was used to make decisions on the complexity of the model, while the final accuracy is always estimated using testing data. The default training and cross-validation datasets consisted of $1000$ noise-only data segments and $1000$ signal + noise data segments at SNR $8$ (i.e. $\rho_{opt}=8$ according to eq.~(\ref{eq:rho_opt})). Finally, for the testing datasets, SNR was varied from $6$ to $14$ in integer steps.

\subsection{\label{ssec:data_features}Data features: speech recognition}

\newcommand{\DEFINEnumberofwavelets}{9}
\newcommand{\DEFINEnumberoffeatures}{184}
\newcommand{\DEFINEnumberoffeaturestotal}{1840}
\newcommand{\DEFINEnumberoffeaturesselected}{4}

\begin{table}[]
	\caption{The speech-processing \& audio analysis libraries used for the feature extraction. For a matter of simplicity, only the most relevant physical and perceptual features from each library are listed in this table. A default set of $\DEFINEnumberoffeatures$ features is extracted from the original timeseries. Then, the latter is pre-processed through a wavelet decomposition, which returns $\DEFINEnumberofwavelets$ pre-processed versions of the waveform. For each pre-processed timeseries, the $\DEFINEnumberoffeatures$ features are again extracted, which rises the total number of features to $\DEFINEnumberoffeatures \cdot (1+\DEFINEnumberofwavelets) = \DEFINEnumberoffeaturestotal$.}\label{tab:features_general}
	\begin{tabular}{@{}ll@{}}
		\br
		\multicolumn{1}{c}{Library} & \multicolumn{1}{c}{Most relevant features} \\ \hline
		\multicolumn{1}{l|}{\href{https://www.idiap.ch/software/bob/}{bob}} & Cepstral coefficients: MFCC, LFCC, RFCC; SCFC, SCMC. \\
		\multicolumn{1}{l|}{\href{https://python-speech-features.readthedocs.io/en/latest/}{python speech features}} & MFCC, LOGF. \\
		\multicolumn{1}{l|}{\href{https://librosa.org/doc/main/feature.html}{Librosa}} & Chromagram features, mel-scaled spectrogram,\\
        \multicolumn{1}{l|}{} & roll-off frequency, rms, various spectral features. \\
		\multicolumn{1}{l|}{\href{https://github.com/tyiannak/pyAudioAnalysis}{pyAudio Analysis}} & Zero crossing rate, energy, entropy, spectral centroid, \\
        \multicolumn{1}{l|}{} &  spectral flux, spectral rolloff, chroma vector.\\
		\multicolumn{1}{l|}{\href{https://essentia.upf.edu/}{Essentia}} & Barkbands, dynamic complexity, hfc, pitch, silence rate,\\
        \multicolumn{1}{l|}{} & various spectral features. \\ \hline \hline
		\multicolumn{1}{c}{} & \multicolumn{1}{c}{Wavelets used for pre-processing stage} \\ \hline 
		\multicolumn{1}{l|}{\href{https://pywavelets.readthedocs.io/en/latest/}{pywt}} & bior, coif9, coif14, sym2, sym17, sym20, db4, db17, db32 \\
		\br
	\end{tabular}
\end{table}

The frequency band of interest for GW searches with second generation interferometers ranges usually from about $10$ Hz to several hundreds Hz. This is in good superposition with the typical frequency band explored in audio signal analyses. For example, the evolution of a GW transient has often been compared to a bird \emph{chirp}. The general idea of our framework is thus inspired by one-dimensional signal processing techniques usually applied for the task of \emph{speech recognition}, i.e. the analysis of audio signals for the purpose of recognizing voice from a persistent background noise. The underlying techniques are well-developed in the literature and are particularly powerful to classify \emph{voiced} signals even in the presence of a particularly \emph{loud} background. The mechanism behind the framework is that of processing data samples through smaller miniframes. In the field of audio analysis and speech-recognition, it is common use to consider the sound on a short-term basis, in order to extract information from the whole audio sequence. Therefore, an audio signal is usually divided into small segments and then the short-term processing stage is carried out. The short-term processing consists in the extraction of physical and perceptual features from each segment. Physical features are measurements computed directly from the sound wave, such as spectrum, energy, entropy, cepstral coefficients, zero crossing rate, and so on \cite{giannakopoulos2015pyaudioanalysis}. Perceptual features are values related to the perception of sounds by the human hearing, such as rhythm, pitch, timbre and noisiness. The feature extraction provides the numerical values that describe properties of an audio segment, so the real information is derived from the analysis of such values, depending on the desired task.

The literature written upon speech processing and audio analysis algorithms is well established \cite{Rabiner10}, and many open-source libraries are already available for use. In this paper, we use a number of existing libraries for speech recognition and audio analysis to compute the short term features used to classify GW interferometer data segments. A full mathematical description of each feature used to derive the ground model is beyond the scope of this paper. Instead, a short resume of the most important libraries and features used for the ground model is reported in Table \ref{tab:features_general}. The total number of features extracted from the libraries is $\DEFINEnumberoffeatures$. In addition to the feature extraction process on the original, whitened, time series, we also added a pre-processing stage for the input waveform, which uses a wavelet decomposition as a low-pass filter. To this end, the wavelet decomposition of a signal is used to partially filter the original waveform. The reconstructed waveform depends on the specific \emph{wavelet} used at first instance, and the spectrum of the reconstructed signal, consequently, is different from the original one. It usually appears smoothed for higher frequencies, while the most energetic tracks in the low-frequency band are highlighted from the surrounding noise. Because we used $\DEFINEnumberofwavelets$ wavelets, $\DEFINEnumberofwavelets$ different pre-processed versions of the timeseries are considered. After filtering the timeseries with the wavelet filter, the basic set of features is again extracted for each pre-processed waveform. The total number of features for each miniframe finally was $\DEFINEnumberoffeatures + \DEFINEnumberoffeatures \cdot \DEFINEnumberofwavelets = \DEFINEnumberoffeaturestotal$.

In terms of performance, the full feature set extraction on $2$k seconds of signal took $240$ seconds on average (the ratio is greater than $8$:$1$), using a commercial $64$-core Threadripper CPU and a simple parallel processing workaround meant to take advantage of the large number of cores. It is important to stress that this result was obtained without a dedicated (low-level) optimization of the feature extraction codes. This raw comparison suggests that the audio feature extraction, upon which the whole framework relies, can ideally keep up to the data buffer required in a real-time situation. In addition, after the model is derived, it is no longer needed to compute all the features described in this section, because the ground model relies on a strongly reduced subset of features, as it will be discussed in detail in Sect.~\ref{ssec:analytical_model}. In fact, one of the crucial aspects of the method used to derive the ground model in the present implementation is the so-called \emph{feature selection}. Feature selection is the capability of a ML algorithm to suitably select a subset of features, among those used for model training, whose informative content is not significantly enhanced when additional features are included. Consequently, one can derive a model that exploits only a reduced number of features, even without significantly affecting the overall accuracy. Usually, a suitable trade-off between complexity (i.e. number of features or their computational complexity) and accuracy is selected. This gives the unique opportunity to explore a large dimensionality feature space, which includes a large body of features derived from speech processing literature, without enhancing the complexity of the resulting model. The latter is a requirement for the ground model of a real time detection algorithm for gravitational waves. To this end, the ground model proposed in this paper is derived using the Brain Project, a state-of-the-art tool for the formal modeling of data based on a hybridization of genetic programming and artificial neural networks \cite{Russo16,Russo20}. This software is particularly suitable for the task of deriving a model aimed to detect GWs in real-time because it has outstanding capabilities of feature selection, as proven for example in Refs.~\cite{Russo14,Buccheri21}.

\subsection{The Brain Project}
\label{sub:bp}
The Brain Project (BP) is a novel tool for the formal modeling of data based on a hybridization of genetic programming and artificial neural networks \cite{Russo16,Russo20}. Genetic programming falls in the field of Evolutionary Computing (EC). In computer science, EC is a branch of artificial intelligence that often deals with optimization problems through algorithms inspired by the Darwinian evolutionary theory \cite{Koza92}. Within this field, BP deals with the so-called symbolic regression, i.e. the process during which some data are fitted by means of a suitable analytical expression derived by the algorithm itself. Many novel algorithms for symbolic regression tasks have recently shown that genetic programming is a particularly powerful technique to solve even extremely complex problems \cite{DellAquila23}. Furthermore, some genetic programming implementations are also capable of feature selection, i.e. to suitably exclude part of the input variables, thus reducing the dimensionality of the feature space.

In the framework of evolutionary computation, a solution of a given optimization problem is initially encoded according to a predefined scheme. Each solution is usually called \emph{individual}, while a set of individuals forms a \emph{population}. A numerical value, called \emph{fitness}, is then associated to each individual in the population. The fitness function quantifies how much a solution is a good solution for the target problem. For instance, in the problem of symbolic regression, i.e. in the task to derive a model for the description of some data, the fitness function might account for the prediction error of the model and/or for its computational complexity. The average fitness in the population is usually maximized (or either minimized, depending on the definition of the fitness function) applying some suitable evolutionary criteria, inspired by the natural selection.

As previously mentioned, in the implementation of BP, genetic programming and NNs cooperate with the aim of deriving a proper analytical model $g({\bf x})$ to fit a set of previously labeled input patterns $\{({\bf x}_i, {\bf y}_i)\ : 1\leq i\leq N_p\}$. The evolutionary computing approach followed in BP is part of a line of genetic programming research that foresees the evolution of tree structures representing formal mathematical expressions. Roughly speaking, to maximize the fitness function, the following steps are executed:
\begin{enumerate}
	\item an initial population of randomly constructed trees, representing solutions of the modeling problem, is generated;
	\item each tree is evaluated through the fitness function and a fitness value is assigned to all individuals in the population;
	\item a new population of trees is created, starting from the population at the previous step, through some evolutionary operators, e.g. copy, crossover, mutation, selection and heuristic operators;
	\item return to step 2, until the termination criterion is met.
\end{enumerate}
In the end, the best individual, i.e. the tree with the maximum fitness, is returned as the output result.

The evolutionary computing core of BP operates a \emph{global} search for the optimal solution of the modeling problem. Such a global search is extremely powerful to identify a solution close to a global maximum of the fitness function, but is usually rather slow to converge to the maximum itself. NNs are therefore used to serve as a hill-climbing operator, i.e. an operator capable to perform a fast local search for the maximum of the fitness function by suitably varying the constants in the mathematical formula. This is a quite time consuming task, thus it is executed only if an individual is particularly promising (i.e. possibly close to a maximum of the fitness function). In the neural part of BP, each expression is initially transformed into a multi-layer feed-forward NN by replacing operators in the original expression with specialized neurons, while all constants are treated as the weighs of the NNs. The mathematical expression of the error gradient in the weight space is then formally calculated and an enhanced gradient descending technique is used to update the weights. Learning weights are usually different for each parameter to optimize and are dynamically varied during the training.

The fitness function used in BP takes into account the prediction error and the number of features exploited by the model (see \cite{Russo16} for additional details).

The software is written in highly optimized C and runs on Linux 64bit environments. It has also a parallel and distributed implementation. Each learning task is executed by more clients controlled by a process called \emph{brain\_server} and running on a server machine. Moreover, BP has also the capability, through a dedicated evolutionary computing algorithm, to self-adapt almost all the internal parameters (\emph{hyperparameters}) required for a learning task. This makes BP a double evolutionary tool \cite{Russo20}, capable of automatically self-tuning itself for the particular problem to solve, thus optimizing the overall performance through, each time, a dedicated set of hyperparameters. BP has been successfully used so far in a number of research fields, see e.g. Refs.~\cite{Russo14,Campobello20,Buccheri21,Buccheri22}, but this is the first time it is used in the field of gravitational wave physics.

\subsection{\label{ssec:analytical_model}Analytical formulation of the model}

Let us summarize the main points discussed so far. The detection of GW signals is based on the analysis of noisy timeseries, in which the dominant frequency range is typically in the audible band. Supported by this fact, the proposed framework is inspired by audio analysis techniques, and specifically by speech processing. One of the key aspects of the newly proposed framework is its layered implementation, which allows to easily specialize the framework for the desired real-time analysis pipeline, even combining different existing algorithms. The lowest level of the framework relies on the initial predictions operated by the ground model. The latter are performed for each individual miniframe, in which the data segment is subdivided, and the successive local analysis (level 2) is performed by analyzing the ground model output for a continuous stream of miniframes. In this paper, we propose a level 1 module whose \emph{ground model} is derived through BP. BP implements a few key advantages for its use in the derivation of real-time GW search algorithms:
\begin{itemize}
	\item BP is capable of a particularly advanced feature selection. For example, the ground model discussed in this paper was initially derived exploiting a full set of $\DEFINEnumberoffeaturestotal$ audio analysis features, but, at the end of the training process, it foresees exclusively $\DEFINEnumberoffeaturesselected$ features. The strongly reduced number of features allows to minimize the computational complexity of the model, thus making it ideal for real-time applications, even if the model is derived using a large body of information on a high-dimensional feature space. In addition, feature selection can also serve as a guideline towards the correct choice of features suitable for the study of GW signals, a crucial information even for the derivation of future algorithms.
	\item The resulting model has a rather simple analytical expression based on numerical features, thus enabling a fast and easy implementation of the model even in low-cost systems based on commercial CPUs.
\end{itemize}

In the mechanism described in Section~\ref{sec:framework}, every data sample of $1$ second is processed in $19$ miniframes of $100$ ms length each, as described in Section \ref{ssec:data_segments}. Given that a typical learning dataset contained $2000$ data samples\footnote{During some preliminary tests, we observe a saturation of the model performance when about $1000$ distinct data samples are used to train the model. For this reason, we conservatively used $2000$ data samples during the training of the ground model.}, BP had in turn $N_p = 19 \cdot 2000 = 38k$ \emph{patterns} available for the training and cross-validation tasks. Each pattern consisted of $\DEFINEnumberoffeaturestotal$ inputs (the features $\{x_i\}$) and a categorical target output value, either $0$, for miniframes containing noise only, or $1$, if the miniframe is in a region where a GW signal is present. For a similar problem, in a high-dimensionality feature space in the presence of large noise, BP usually achieved a convergence within a time of a few hours using a single Threadripper CPU. With a similar procedure as described for example in Refs.~\cite{Campobello20, Buccheri21}, after a few preliminary runs, we settled to an optimal trade-off between complexity and accuracy. An analogous procedure was also applied to set the optimal miniframe length to $100$ ms; in this respect, we performed preliminary learning runs and tested the deriving preliminary models on an additional subset of the cross-validation data. The performance of the model, for each selected trade-off, was monitored by analyzing the prediction error on the cross-validation dataset, which contained the same number of patterns as in the training dataset.

The analytical implementation of the ultimately derived model is reported in Table~\ref{tab:formulation}, together with a brief description of the corresponding numerical features. Only $\DEFINEnumberoffeaturesselected$ features are exploited by the model, and it is therefore not needed to compute all the features described in Table~\ref{tab:features_general}. The output value, $y_i = f(x^i_1, ..., x^i_{\DEFINEnumberoffeaturestotal})$ is usually not defined in the range [$0$, $1$]. In order to fulfill the condition $f({\bf x})\in[0,1]$ for all possible input vectors ${\bf x}$, it is convenient to use the corresponding constrained function $g({\bf x})=\max(\min(f({\bf x}), 1),0)$.
\begin{table*}[t]
	\caption{The analytical formulation ultimately derived by BP for the miniframe output value $y$. Despite the complexity of the problem and the large dimensionality of the feature space used to train the model, the derived formula foresees exclusively $\DEFINEnumberoffeaturesselected$ features and has a particularly low computational complexity. This model is used to perform a ground prediction on each miniframe, individually.}\label{tab:formulation}
	\begin{tabular}{@{}ll@{}}
		\br
		
		\multicolumn{2}{c}{ $ y = - 0.07133 + 0.86274 \cdot pyAA_{31} \cdot \chi + 70.47328 \cdot pyAA_{33}$ } \\ \multicolumn{2}{c}{  where $\quad  c = {\max\left ({0.22473}+{{\ln\left(db17_{ess3}\right)}\cdot {\erf\left(db32_{bb4}\right)}}-{{2.6911}\cdot {pyAA_{31}}}, \; db17_{ess3} \right)}$ } \\ 
        \multicolumn{2}{c}{ $\chi = \exp \left[ \min \left( 0.036685, db17_{ess3} \right) - {1.0623 \cdot {db17_{ess3}}} + c \; \right]$ }
        \\ \hline
		
		\multicolumn{2}{l}{the following $\DEFINEnumberoffeaturesselected$ features have been selected:} \\
		\multicolumn{1}{l|}{$pyAA_{31}$} & The second last element of the \emph{chroma vector}\tablefootnote{The chroma vector is a 12-element representation of the spectral energy, where the bins represent the 12 equal-tempered pitch classes of music (in the western-type semitone spacing).} given by the pyAA library.  \\
		\multicolumn{1}{l|}{$pyAA_{33}$} & The standard deviation of the 12 elements of the chroma vector\\
        \multicolumn{1}{l|}{} & given by the pyAA library. \\
		\multicolumn{1}{l|}{$db17_{ess3}$} & The energy of the waveform in band ]150,800] Hz given by Essentia library\\
        \multicolumn{1}{l|}{} & on the db17 pre-processed waveform. \\ 
		\multicolumn{1}{l|}{$db32_{bb4}$} & The energy of the waveform in band ]200,300] Hz given by Essentia library\\
        \multicolumn{1}{l|}{} & on the db32 pre-processed waveform.\\
		\br
	\end{tabular}
\end{table*}

\section{\label{sec:results}Results and discussion}

\subsection{Interpretation of model output: multi-modeling in the task of local analysis}
\label{subsec:level2}
One of the key aspects of the proposed method is that the derived model has an analytical formulation, entailing only a few nodes. This allows to perform a prediction with a reduced computational complexity compared to CNN approaches. In addition, the layered framework requires suitable interpretation at each level of decision, making the framework particularly versatile for many real-time applications.

From a detailed inspection of the ground model equation of Table~\ref{tab:formulation}, one can point out some interesting points. The features exploited by the model are exclusively: $pyAA_{31}$, $pyAA_{33}$, $db17_{ess3}$, $db32_{bb4}$. This indicates that their informative content, for the task of identifying GW signals in noisy and short time intervals, is equivalent to the informative content of the $\DEFINEnumberoffeaturestotal$ features initially selected for the learning task and reported in Table~\ref{tab:features_general}. In other words, forcing the model to exploit all the features of Table~\ref{tab:features_general} would increase the complexity of the model without a statistically significant reduction of the prediction error. Selected features are briefly described in Table~\ref{tab:formulation}. The first two features, extracted from the library $pyAA$, are related, respectively, to the energy of the strain in the frequency band from about $98$ Hz to about $415$ Hz, and to the overall variation of energy (expressed as the standard deviation) in the frequency domain from $30$ Hz to $1000$ Hz. The latter might usually exhibit lower values if the miniframe is strongly dominated by noise, as the noise power is equal at all frequencies after whitening the data segment, but will tend towards larger values if the miniframe contains a GW signal, whose power lies, for a given miniframe, in a narrow frequency region. The second two features are calculated from the signal initially filtered through the Wavelet filter. They represent, respectively, the energy of the pre-processed strain in the band $]150,800]$ Hz and $]200,300]$ Hz. Interestingly, selected features are related exclusively to the energetic of the strain, suggesting that energy is the most relevant information for the detection of GWs in a short-term feature approach. Furthermore, it is not surprising that all selected features are linked to a frequency range that contains the peak amplitude of BBH waveforms, in the mass range explored in the present paper.

Unlike previous classifiers based on CNNs, the ground model itself does not provide a final prediction on a fixed time length segment, because it is strictly trained on local properties of the signal. For this reason, the generation of an interferometer alert requires additional dedicated layers for the refined interpretation of the ground model information. Usually, the algorithm can be suitably adapted to the desired detection task, eventually by reaching a given trade-off between sensitivity and specificity of the classification. For example, lower levels should be extremely conservative in ruling out noise segments, in order to obtain the adequate level of sensitivity required for the successive offline analysis and parameter estimation.

\begin{figure}[t]
	\centering
	\includegraphics[width=0.45\textwidth]{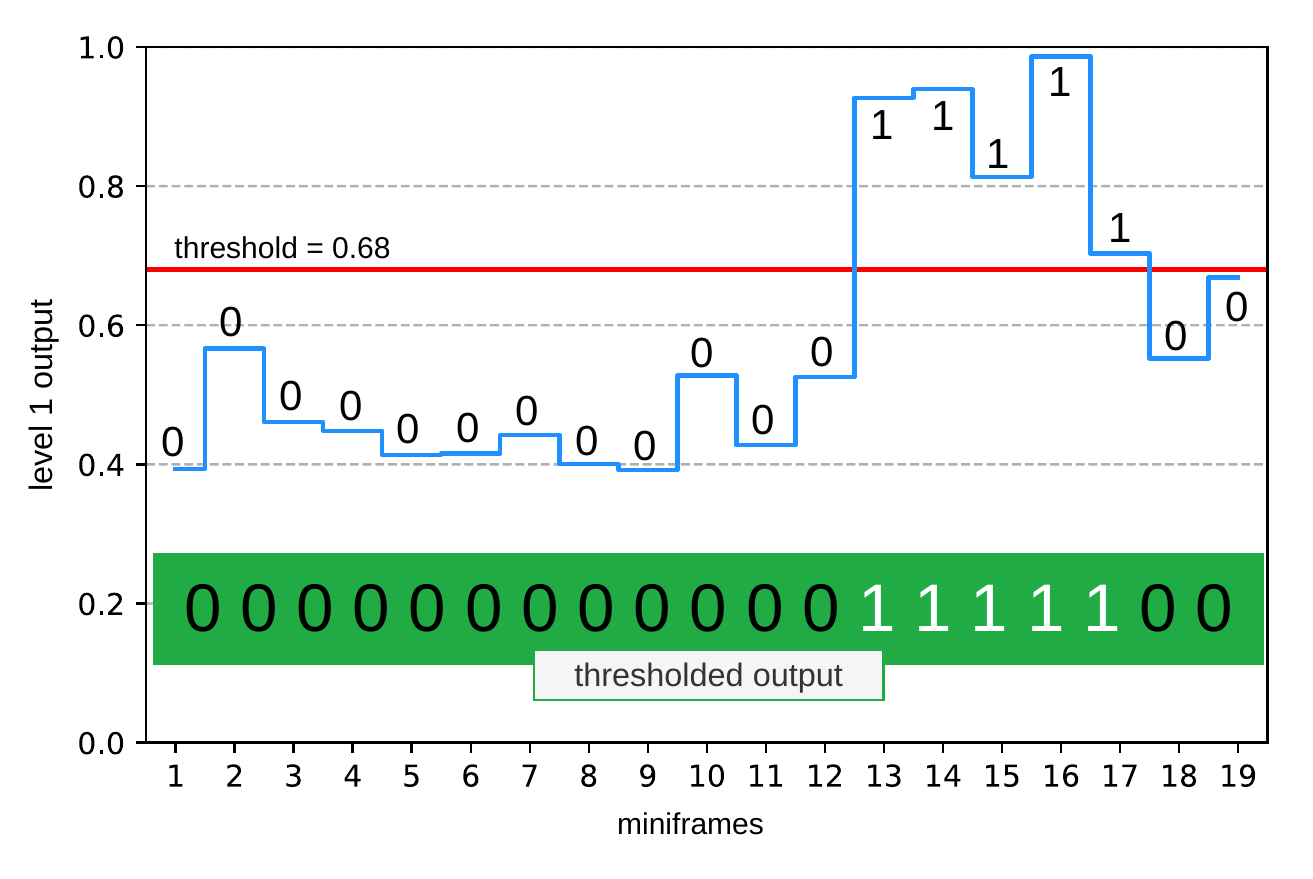}
	\caption{\label{pic:threshold_positive} A stream of $n$ ground model output values $y_i$ for $19$ consecutive miniframes of $100$ ms length. Data are produced using a signal+noise data segment extracted from the testing dataset. The total duration of the segment is $1$ second, thus the miniframes are overlapped by $50 \%$ of their duration. Model output clearly exhibits an increasing trend from left to right, reflecting the peak amplitude of the BBH waveform. By thresholding the output values with a fixed threshold, one obtains a categorical output. The latter exhibits a cluster of positive output (i.e. $1$) at the peak amplitude. This information is suitable to local analysis of the segment and therefore the formation of level 2 output.}
\end{figure}
\begin{figure}[t]
	\centering
	\includegraphics[width=0.45\textwidth]{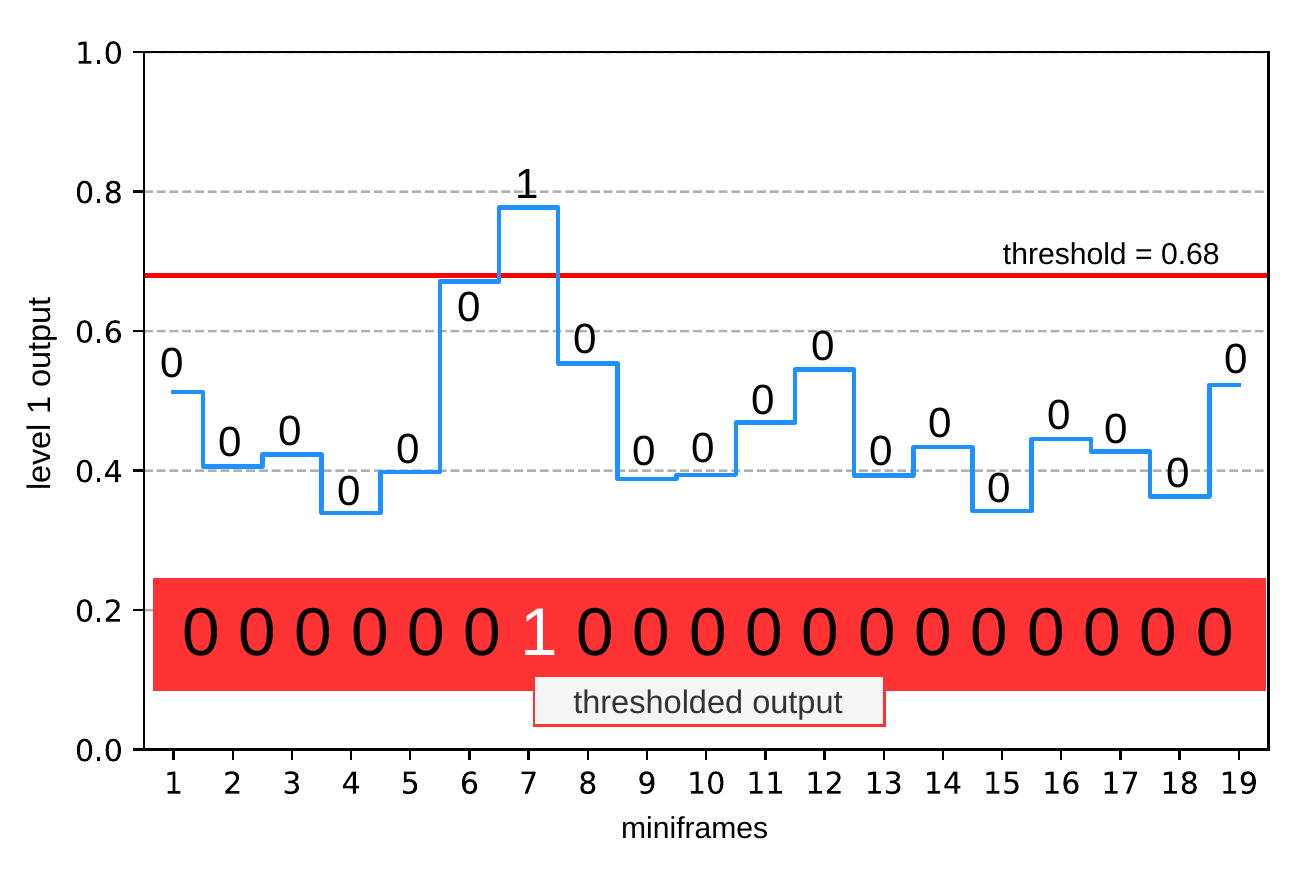}
	\caption{\label{pic:threshold_negative} Same as Figure~\ref{pic:threshold_positive} but obtained with a noise only sample. In this case, the output has lower variation and a lower mean-value. The thresholded categorical output does not exhibit clusters of $1$ and the $0$ prediction dominates the data segment.}
\end{figure}

In the proposed framework, regardless of the target task, either local or continuous analysis, a generic data segment from a GW interferometer is considered through $n$ overlapped miniframes of fixed time length ($\omega_f = 100$ ms). For each miniframe, the reduced set of speech-processing features is extracted and used to compute a continuous output $y$ (Table~\ref{tab:formulation}), suitably constrained in the range $[0,1]$ as indicated in Section~\ref{ssec:analytical_model}. At this stage, the generic segment is matched to a stream of $n$ consecutive outputs $y_i$. In this paper, we mostly focus on the local analysis, to have a more meaningful comparison with CNN approaches \cite{Gabbard18,George18,George18b}. A typical stream of ground model output for a locally analyzed $1$-second frame is shown in Figure~\ref{pic:threshold_positive}, for a signal+noise segment, and in Figure~\ref{pic:threshold_negative} for a noise only segment. Both data segments used to produce the figures are extracted from the testing dataset, and are therefore good representative of the typical level 1 output. By inspecting the figures, it is clear that the observed trends for the model output, as a function of the miniframe, differ significantly for signal+noise and noise only segments. The first, Figure~\ref{pic:threshold_positive}, exhibits an increasing trend, with a group of consecutive miniframes having a much larger output. The latter has instead a more uniform output, reflecting the roughly uniform behavior of the detector noise in a short data segment.

\subsubsection{L2T: a possible level-2 implementation based on a thresholded model}
A first possible method for the interpretation of the ground model output, for the task of local analysis (performed by level 2), can be obtained by suitably thresholding its output. The principle is schematically explained in Figures~\ref{pic:threshold_positive} and ~\ref{pic:threshold_negative}, where a threshold of $\tau=0.68$ is represented, as an example. After thresholding the output, one obtains a series of $19$ categorical output values for each data segment to analyze (with values either $0$ or $1$). Following the different trend of model output for the two cases, one expects the appearance of clusters of $1$ in the presence of a GW signal, while for noise only segments only some, spurious, isolated $1$ is usually present, as seen in Figures~\ref{pic:threshold_positive} and \ref{pic:threshold_negative}. One can exploit this difference in the observed patterns of the categorical output to derive a model for the \emph{thresholded} interpretation of the ground model. For the thresholded interpretation model, we used maximum length of a cluster of $1$, the maximum length of a cluster of $0$ and the total number of $1$ observed in the segment. Boundary conditions on these three variables can be derived through any ML algorithm, exploiting a dataset of previously labeled segments. We name this approach \emph{multi-modeling}. To this end, we exploited $1$-second data segments, previously simulated, and subdivided into training, cross-validation and testing datasets, dedicated to the derivation of the interpretation models.

\begin{figure}[t]
	\centering
	\includegraphics[width=0.5\textwidth]{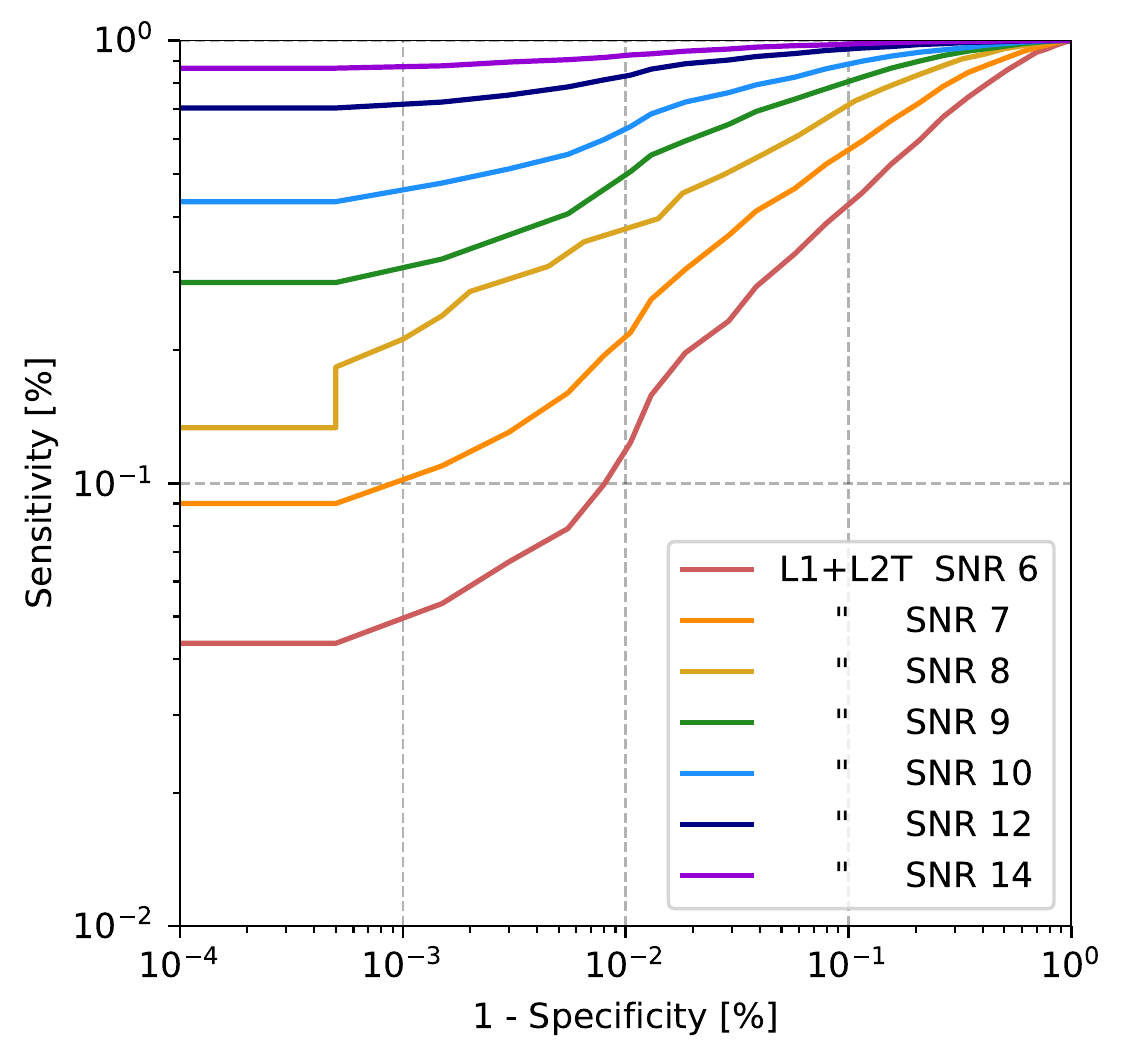}
	\caption{\label{pic:roc_threshold_log} ROC curves obtained by a local analysis of the testing dataset with L1+L2T. The Y-axis represents the sensitivity, i.e. the fraction of data segments containing a GW signal triggering the second layer of the framework, and the X-axis represents the false alarm probability, i.e. the fraction of noise-only segments incorrectly producing a trigger. The scales are logarithmic. Curves are shown, individually, for optimal SNR ranging from $\rho_{opt}=6$ to $\rho_{opt}=14$ with different colors. One observes, as expected, a raise in performance of the model towards higher SNRs.}
\end{figure}
Using the derived boundary conditions on the above mentioned quantities, the performance of the thresholded version of level 2 (L2T) can be evaluated using the segments of the dedicated testing dataset. Figure~\ref{pic:roc_threshold_log} shows the receiving operator characteristic (ROC) curves obtained using L1 and the thresholded version of L2 (L1+L2T) for optimal SNR ranging from $\rho_{opt}=6$ to $\rho_{opt}=14$, individually. In the graph, the Y-axis is the specificity of the classifier, i.e. the fraction of signal+noise segments that are correctly classified as containing a GW signal. This is a crucial quantity in real-time GW detection applications. X-axis is related to the capability of the model to rule out noisy segments. It corresponds to the fraction of noise-only samples incorrectly classified as containing a GW signal. For the latter, $0$ indicates a complete, ideal, noise rejection, while $1$ indicates that all noise-only segments in the dataset produce a zero-level trigger. Each point in the ROC curve is obtained with a certain value of the threshold $\tau$. A threshold value $\tau=1$ would result in a point located at the bottom-right corner of the graph, i.e. all segments are identified as noise, while $\tau=0$ will identify all segments as containing a GW, top-right corner. An optimal LT2 can be obtained selecting the proper threshold, according to the required performance. For example, for an optimal SNR of $\rho_{opt}=12$, one can easily rule out $90 \%$ of the noise-only sample, while identifying nearly-$100 \%$ of the GW signals. This value drops to about $70 \%$ for $\rho_{opt}=8$. In general, L1+L2T exhibits a good regularity for increasing SNRs, with a significant rise in performance towards higher SNR values. Furthermore, the best trade-off between specificity and sensitivity is obtained at a similar threshold value at all SNR, consistently.

\begin{table}[t]
	\caption{The analytical formulation of the unthresholded model derived via BP. A brief description of the selected features is also provided. The unthresholded model foresees only $6$ features. See the appendix for a detailed description of the features selected by BP.}\label{tab:features_unthresholded}
 \centering
	\begin{tabular}{lll}
    \br
        \multicolumn{3}{c}{$ Y = \tanh \Big\{  1.83 \cdot \max \left[ \;  c_1  , \; \max( c_2, b_4) \; \right] \Big\} ^{19.5073\cdot \arctan(b_6)}$} \vspace{3mm}  \\
        \multicolumn{3}{c}{where $ c_1 = 1.829 \cdot b_4 \cdot \max\Big\{ 1.829 \cdot b_4  \cdot \tan(b_4)  , \;\;   \xi + 0.947062 \cdot b_2  \Big\} + b_2 $,}\\
        \multicolumn{3}{l}{ \hspace{4em} $\xi = 1.732176  \cdot b_4 \cdot \sin [ \max( \tan(b_3) , b_4 ) ]$}\\ 
        \multicolumn{3}{l}{ \hspace{4em} and $ c_2 = \{ 1.829 \cdot \sinh [  \erf  (  b_3 \cdot b_1 )  ] \cdot \erf [ \tan(b_4) ] \cdot b_5  \}  + b_2 $ } \\[3mm] 
        \hline
        
        \multirow{3}{*}{The features:} & $b_1$ = $russo8_{3}$ & $b_2$ = $russo9_{5}$ \\
                               & $b_3$ = $sm2ampl$ & $b_4$ = $sm3weav$ \\
                               & $b_5$ = $sm5barone7_{2}$ & $b_6$ = $sm5barone9_{0}$ \\
		

		\br
	\end{tabular}
\end{table}

\subsubsection{L2U: an alternative level-2 implementation with an unthresholded model}
\begin{figure}
	\centering
	\includegraphics[width=0.5\textwidth]{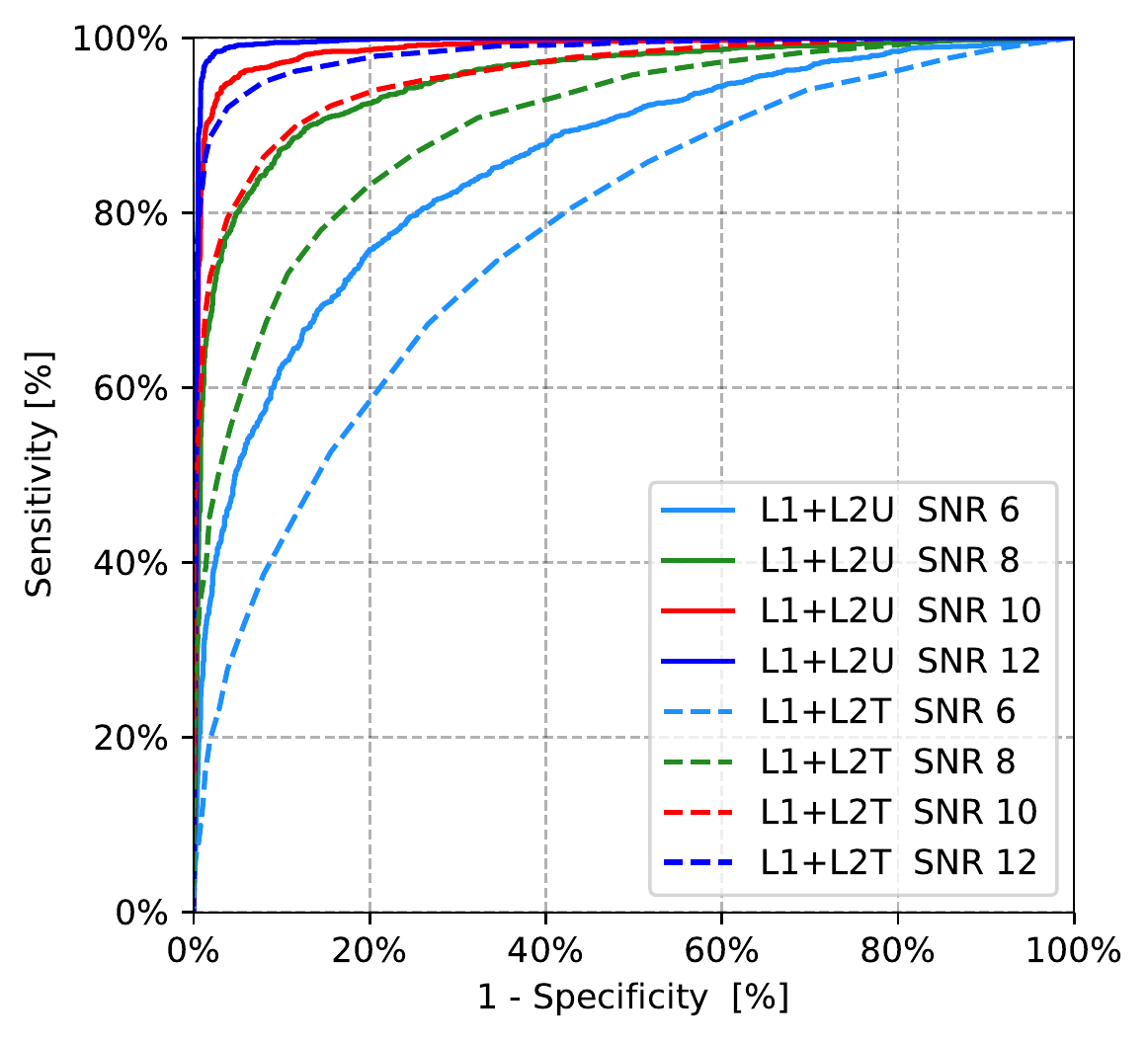}
	\caption{\label{pic:roc_unthresholded} Local analysis ROC curves obtained with L1+L2U (solid lines) compared to those of L1+L2T (dashed lines) at analogous SNR. Each color represents a given SNR. Optimal SNR values were varied from $\rho_{opt}=6$ to $\rho_{opt}=12$. As in Figure~\ref{pic:roc_threshold_log}, Y-axis represents the sensitivity of the classifier and X-axis is the false alarm probability. L2U performs better than L2T at all SNRs.}
\end{figure}

An alternative way to interpret the ground model output can be achieved without thresholding the values of level 1. In fact, thresholding the output typically results in a loss of information, being the continuous output converted into a categorical one with only two possible categories. To derive the unthresholded model, we considered the ground model output values for $19$ consecutive miniframes in which the data segment is subdivided. Said $y_i$ the output of the $i$-th miniframe, we constructed the curve identified by the set of points $\{(i,y_i) : i\in[1,19]\}$. To derive the present version of the L2U model, we decided to build a set of $260$ features to represent the information on the presence of a GW signal in the noisy segments. Such features included simple statistical values (mean, median, standard deviation, kurtosis) and linear fit coefficients. The same set of features is also extracted on multiple smoothed version of the continuous output curve, obtained with different smoothing, as described in \ref{appendixb}. Exploiting this set of features, a second model was derived through BP, using the data segments contained in the same subset of the testing dataset previously used to derive the thresholded model. The deriving model can be used to build an alternative unthresholded level 2 (L2U). The analytical implementation of the unthresholded model is shown in Table~\ref{tab:features_unthresholded}, together with a description of the selected features. A more detailed description of the features used by the unthresholded model is instead reported in the appendix. Similarly to the case of the ground model, also the unthresholded model exploits an extremely reduced number of features. Features are reduced from the initial $266$, in the training set, to only $6$. The performance of the deriving L2U in the task of identifying GW data segments is probed using the same dataset previously used to test the performance of L2T. This time, a given trade-off between sensitivity and false positive rate can be obtained by using a threshold $\eta$ directly on the output of the unthresholded model. The corresponding ROC curves, in which each point corresponds to a given $\eta$ value, are shown in Figure~\ref{pic:roc_unthresholded} (solid lines), compared to those of L1+L2T (dashed lines). Curves obtained at analogous SNR are drawn with the same color.

An inspection of the ROC curves for the two modules suggests that the performance of L1+L2U is significantly better than that obtained with L1+L2T (and therefore by thresholding the ground model output). As an example, for an optimal SNR $\rho_{opt}=8$ and a false positive rate of $10 \%$, the sensitivity of L1+L2U raises to about $90 \%$, while L1+L2T has a sensitivity of only $70 \%$. Interestingly, for optimal SNR $\rho_{opt}=12$, the sensitivity of L1+L2U is almost saturated to $100 \%$ at nearly all false alarm probability values, while, using the thresholded module leads to a sensitivity below $90 \%$ when the false alarm probability approaches $5 \%$. This indicates that the new set of features used for the L2U model has a much larger informative content than the simple quantities used in the L2T model. It is clear that other alternative interpretations of the ground model output (and therefore other possible level 2 implementations), other than the ones proposed here, could lead to even more accurate predictions.

\subsection{\label{ssec:comparison_dnn} Comparison with the literature}
The local analysis described in the previous section, i.e. the analysis of data segments of fixed length, allows a direct comparison of L1+L2 derived in the present work with other state-of-the-art approaches available in the literature. We compare the results achieved with L1+L2U to those of matched-filtering and state-of-the-art artificial intelligence approaches based on CNNs previously exploited in GW detection. We consider the CNN model described in Ref.~\cite{Gabbard18}. This model is derived and tested exclusively on data segments of $1$ second length, where the GW peak lies in the fractional range [$0.75$, $0.95$] of the time series. For this reason, it is ideal to be used for the local analysis proposed in this section. Matched-filtering is a key algorithm, usually adopted for the detection of binary coalescence signals \cite{Babak13, Allen12}, which is based on the coherence of the detected data segment with a given template waveform. The ranking statistics used for matched-filtering is the matched-filtering SNR numerically maximized over arrival time, phase and distance (see Ref.~\cite{Gabbard18} for additional details). Given the ranking statistic of each analysis algorithm, it is possible to construct the ROC curves by evaluating, for each threshold, the number of true positives and true negatives achieved by the classifier.
\begin{figure}
	\centering
	\includegraphics[width=0.5\textwidth]{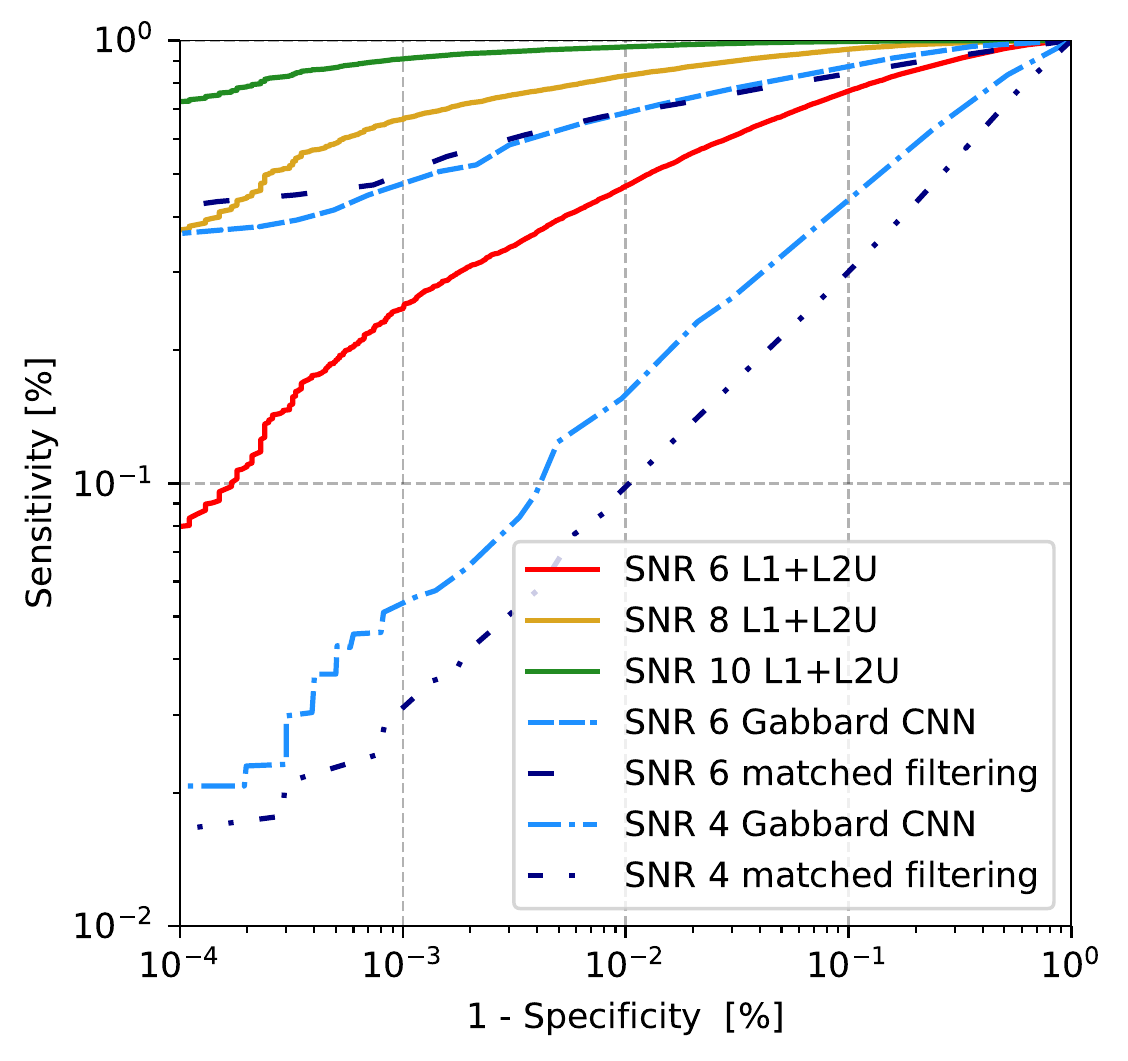}
	\caption{\label{pic:comparison_literature} ROC curves obtained with L1+L2U (solid red curve, $\rho_{opt}=6$, solid yellow curve, $\rho_{opt}=8$, solid green curve, $\rho_{opt}=10$), the standard matched-filtering (dark blue curves at optimal SNR $\rho_{opt}=6$), and the CNN model of Ref.~\cite{Gabbard18} (light blue curves, at optimal SNR $\rho_{opt}=6$). For matched-filtering and the CNN model also $\rho_{opt}=4$ is shown (dash-dotted lines). ROC curves for matched-filtering and the CNN models are extracted from Ref~\cite{Gabbard18}, where an analogous dataset as the one used in the present paper is adopted.}
\end{figure}

In Figure~\ref{pic:comparison_literature}, we show the ROC curves obtained with the present framework using the most accurate module derived in this work, together with the corresponding ROC curves obtained using standard matched-filtering and the model described in Ref.~\cite{Gabbard18}. We focused the comparisons on data containing signals with optimal SNR ranging from $\rho_{opt}=6$ to $\rho_{opt}=10$, while, for matched-filtering and the CNN model, we also show the results for $\rho_{opt}=4$, which is well below the currently considered confident detection limit\footnote{A confident GW detection conventionally requires $\rho_{opt}\geq8$. The signals used to test the capabilities of the framework are thus extremely challenging signals.}. The curves obtained at $\rho_{opt}=6$ with matched-filtering (blue dashed line) and the CNN model (light blue dashed line) show similar values of sensitivity and specificity. The sensitivity of the CNN slightly exceeds that of matched-filtering at large false alarm probability values, while the trend is opposite at small values of false alarm probability, i.e. at large sensitivity values. The ROC curves for L1+L2U are shown with the red, yellow and green curves for $\rho_{opt}=6$, $8$, and $10$, respectively. Compared to the CNN model at the same SNR, CNNs have better performance than our model for all values of the false alarm probability. For example, at a false alarm probability of $10^{-2}$, and for $\rho_{opt}=6$, L1+L2U is able to identify about $45 \%$ of the GW data segments in the testing dataset, while CNNs identify about $65 \%$ of the GW signals. At optimal SNR $\rho_{opt}=4$ (dash-dotted lines), the performance of CNNs drops significantly compared to our results at $\rho_{opt}=6$. If we focus on conventionally considered \emph{confident} SNR values, i.e. $\rho_{opt}\geq8$, the performance of our simple classifier is higher than that of CNNs and matched-filtering at $\rho_{opt}=6$. At $\rho_{opt}=10$, we obtain an almost saturated sensitivity at all values of false alarm probability $\geq10^{-3}$. However, one must consider that the proposed models are outstandingly simple. Finally, it is important to stress that the situation of local analysis, which is a particular ideal benchmark for the CNN model of Ref.~\cite{Gabbard18}, is a particularly unfavorable condition for our framework, which is instead designed for a short-term analysis of features, thus not accounting for the duration of the strain and the position of the signal peak within the segment. The assumptions used to derive our models are ideally more advantageous for the analysis of a continuous timeseries in a realistic analysis pipeline. In addition, as shown previously in the paper, the derived models (for both level 1 and level 2) have the further advantage of an outstandingly-low computational complexity, which makes them ideal for the integration in real-time, low-latency, analysis pipelines.

\section{\label{sec:conclusions}Conclusions and perspectives}
Developing fast and accurate methods for real-time detection of gravitational wave signals is a crucial task in the era of multi-messenger astrophysics. In this foundational paper, we describe a new framework designed for the fuzzy-identification of gravitational waves in real-time, low-latency, applications. The proposed framework foresees a layered approach, in which less computationally complex operations are executed as the first layers, to rule out regions of the interferometer timeseries exclusively characterized by instrumental noise. The framework is constructed around the idea of using features derived from speech processing to identify GW signals in noisy time series. This is a novelty in the research field of gravitational wave physics.

We describe the derivation of a model suitable to serve the first level of the framework, i.e. the only module that directly deals with the inferferometer timeseries. Such level 1 relies on the \emph{ground} predictions of the so-called \emph{ground model}, which performs a short-term analysis of the time series. To this end, the data segment to analyze is suitably subdivided into overlapped miniframes and an extremely reduced set of $\DEFINEnumberoffeaturesselected$ state-of-the-art speech processing features is calculated for each miniframe. The Brain Project, a novel hybridization of genetic programming and neural networks for the formal modeling of data, is used to derive a model capable to link short-term features to the presence of a GW signal. The model is derived using a dataset of simulated BBH signals in synthetic Gaussian noise representative of aLIGO, and an extended set of $\DEFINEnumberoffeaturestotal$ features. One of the novelties of the approach is that the deriving model has an analytical formulation, which can be easily implemented even in systems based on commercial CPUs. Thanks to the outstanding feature selection capabilities of the Brain Project, the ultimately derived model foresees only $\DEFINEnumberoffeaturesselected$ features, thus having an extremely low computational complexity. Results of the ground model are interpreted through the second level of the framework. Two possible versions of such level are proposed, based on two different dedicated interpretation models derived using artificial intelligence supervised learning approaches, namely thresholded and unthresholded models.

The performance of the two alternative implementations of level 2, upon level 1, is probed via a detailed inspection of the receiving operator characteristic curves obtained from the analysis of data segments of fixed length. We used a dataset containing signals at low SNR and a meaningful astrophysical parameter distribution suggested by the literature. Results show that avoiding thresholding the ground model output leads to significantly better results. As an example, for signals with optimal SNR $\rho_{opt}=12$, at a false alarm probability of $10 \%$, the sensitivity (i.e. the true alarm probability) of the unthresholded interpretation is nearly $100 \%$, while the thresholded one identifies only about $95 \%$ of the GWs in the dataset. Results are compared with standard matched-filtering algorithms and a recent CNN model, previously validated on data segments of fixed length. The unthresholded model leads to a performance slightly lower than that of matched-filtering and CNNs. For example, at $\rho_{opt}=6$ and a false alarm probability of $10^{-2}$, CNNs achieve $65 \%$ sensitivity while the unthresholded model allows to reach a sensitivity of $45 \%$. However, it is important to stress that the local analysis of fixed segments is a particularly unfavorable working condition for our framework, which is instead ideal to analyze a continuous buffer of interferometer data, as in a real experiment case.

The paper proves that speech processing features can be used as short-term features for the detection of GWs in noisy time series detected by second generation interferometers, such as aLIGO. The proposed method could help to develop new dedicated pipelines for the real-time analysis of GW interferometer data at extremely low-latency. Furthermore, the high-versatility of the framework and the optimized implementation of the underlying models makes it suitable to be integrated even in existing analysis pipelines. In addition, these results are particularly promising also for future development towards the third generation of gravitational wave interferometers.

\section*{Acknowledgements}
D.D. acknowledges funding support from the Italian Ministry of Education, University and Research (MIUR) through the ``PON Ricerca e Innovazione 2014-2020, Azione I.2 A.I.M., D.D. 407/2018". The authors gratefully acknowledge Dr. H. Gabbard (University of Glasgow) for fruitful discussion. The authors would like to acknowledge the G2NET Cost Action (CA17137) for the valuable support and stimulating discussions.

\section*{Appendix}
\appendix
\section{Simulation of GW signals}
\label{app:gw}
To simulate the GW signals used to produce GW data segments, we exploited the LALsuite library \cite{lalsuite} and adopted the IMRPhenomD-type waveform \cite{Husa16,Khan16}. The characteristic \emph{chirp} of BBH waveforms, comprising inspiral, merger and ringdown phases, is obtained by matching together a high-order 3.5PN waveform for the inspiral \cite{Blanchet98,Blanchet02,Blanchet04} and a numerical waveform for merger and ringdown \cite{Pretorius05, Baker06, Campanelli06}. As proven by \cite{Kidder08}, it is advantageous to project the waveform onto spin-weighted spherical harmonics prior to truncating the post-Newtonian series of the numerical simulation.
Given the asymptotic waveform $h^{TT}_{ij}$ and a orthonormal triad $ (\vec{N},\vec{P} , \vec{Q}) = (\vec{e}_R, \vec{e}_\Theta,\vec{e}_\Phi )$, the polarization waveforms are
\[ h_+ = \frac{1}{2}(P_iP_j-Q_iQ_j)h^{TT}_{ij} \]
\[ h_{\times} = \frac{1}{2}(P_iQ_j+P_jQ_i)h^{TT}_{ij} \]
which can be decomposed using spin-weighted spherical harmonics of weight -2 as
\[ h_+ - ih_{\times} = \sum_{\ell=2}^{\infty} \sum_{m=-\ell}^{\ell} h^{\ell m} _{-2}Y^{\ell m}(\Theta,\Phi) .\]  
The polarization waveforms are then processed by an antenna response function to simulate their detection from an ideal ground-based interferometer, while no background noise is added yet. For the component masses of the binary system, $m_1$ and $m_2$, we assumed a meaningful astrophysical distribution, which is the canonical logarithmic mass distribution described in \cite{Abbott16e}. $m_1$ is randomly extracted from $5$ to $100$ solar masses, so that $m_2 < m_1$ and $m_1 + m_2 < 100 M_\odot$. The astrophysical mass distribution used to generate the waveforms allows not only to probe the capabilities of the model on a \emph{realistic} physics case, but also to perform a detailed direct comparison with other, previously published, state-of-the-art detection algorithms based on artificial intelligence \cite{Gabbard18}. We considered only non-spinning black holes. The other parameters assume an isotropic distribution of the gravitational source in the sky:
\begin{itemize}
	\item inclination angle is generated according to a $\sin$ distribution in the interval [$0$,$\pi$];
	\item phase and polarization angle are generated according to a uniform distribution in the range [$0$, $2\pi$]; 
	\item a random sky location is extracted uniformly on a sphere, and the source is placed at a fixed distance of $1$ Mpc.
\end{itemize}

\section{Selected features of the unthresholded model}
\label{app:unthresholded}
\subsection{Standard definitions}

For each 1 second sample there are 19 continuous miniframe outputs $y_i$ ($i = 1 ... 19$), usually in the range [0,1]. Let us call $\Gamma$ the curve given by the set of points $(i, y_i)$:
\[ \Gamma = \{ (i, y_i) : i = 1 ... 19 \} .\]

The original curve given by the miniframe outputs can be smoothed considering the mean value on the Y-axis of $p$ points, $p>1$.
By definition, a \emph{mean-smoothed curve} $\Gamma_p$ is the set of points $(j,Y_j)$, for $j = 1 ... (19+1-p)$, in which every point $Y_j$ is given by the mean Y-axis value of $p$ adjacent points of $\Gamma$:
\[ Y_j = \frac{1}{p} \sum_{k=0}^{p-1} y_{j+k} ,\]
\[ \Gamma_p = \{ (j, Y_j) : j = 1,...,19+1-p \} .\]
Because of this definition, the not-smoothed curve $\Gamma$ is a smoothed curve $\Gamma_1$ with $p=1$. For a given curve $\Gamma_x$, let us consider $\alpha$ its amplitude, $\alpha_x = \max\Gamma_x - \min\Gamma_x$, and its minimum value $\mu_x = \min\Gamma_x$

Each point of the curve is processed by threshold, so that every point above the threshold value $\tau$ is considered as a 'one', otherwise as a 'zero'. Within each curve, except for the case in which the threshold value is grater than the minimum value of the curve, there will be a 'first one' and a 'last one' in the sequence. Let $\xi$ be the distance between the 'first one' and the 'last one' of the curve on the X-axis.

\subsection{Feature definitions}
\label{appendixb}
Considering $\Gamma$ (not-smoothed curve):
\[ b_1 = \frac{1}{\xi}\sum_{ones} y_i \]
where the ones are assigned using a threshold value $\tau = \mu + 0.8\cdot\alpha$.
\[ b_2 = \sum_{i=0}^{19} \Big( y_i - (0.9 \cdot \alpha + \mu)\Big) \]

Considering a smoothed curve with $p=2$:
\[ b_3 = \alpha_2\]

Considering a smoothed curve with $p=3$:
\[ b_4 = \sum_{j=1}^{17} (Y_j \cdot w_j) \Bigg/ \sum_{j=1}^{17} w_j \]
where $\{w_j\}$ is a suitable weighting vector to account for the enhancement of the signal power from the left to the right part of the data segment, as a result of the topology of the BBH waveform.

Considering a smoothed curve with $p=5$ and $\tau = 0.7\cdot\alpha + \mu$, $b_5$ is the length of the shortest cluster of zeros.

Considering a smoothed curve with $p=5$ and $\tau = 0.7\cdot\alpha + \mu$, $b_6$ is the number of the ones within the curve.

\section*{References}
\bibliographystyle{iopart-num}
\bibliography{bibliography_gw_20210528}

\end{document}